\documentclass[aps,pra,reprint,twocolumn,floatfix,superscriptaddress,longbibliography]{revtex4-1}

\usepackage{standalone}
\usepackage{amsthm}
\usepackage{lipsum}
\usepackage{amsfonts, amsmath, amssymb}
\usepackage{graphicx}
\usepackage{dcolumn}
\usepackage{bm}
\usepackage{dsfont}
\usepackage[normalem]{ulem}
\usepackage{color}
\usepackage[utf8]{inputenc}
\usepackage[colorlinks,citecolor=blue,linkcolor=blue,urlcolor=blue]{hyperref}
\usepackage{tikz}
\usepackage{braket}
\usepackage{physics}
\usepackage{color}
\usepackage{soul}
\usepackage{mathtools}
\usepackage{float}
\usepackage{esvect}
\usepackage{subfigure}
\usepackage{stmaryrd} 
\usepackage{enumerate}

\include{physics}
\begin{document}

\title{Metrics and properties of optimal gauges in multimode cavity QED}

\author{Geva Arwas}
\affiliation{Universit\'{e} Paris Cit\'e, CNRS, Mat\'{e}riaux et Ph\'{e}nom\`{e}nes Quantiques, 75013 Paris, France}

\author{Vladimir E. Manucharyan}
\affiliation{\'{Ec}ole Polyt\'{e}chnique F\'{e}d\'{e}rale de Lausanne, 1015 Lausanne, Switzerland}
\author{ Cristiano Ciuti}
\affiliation{Universit\'{e} Paris Cit\'e, CNRS, Mat\'{e}riaux et Ph\'{e}nom\`{e}nes Quantiques, 75013 Paris, France}

\begin{abstract}
\noindent
Multi-mode cavity quantum electrodynamics (QED) describes, for example, the coupling between an atom and a multi-mode electromagnetic resonator. The gauge choice is important for practical calculations in truncated Hilbert spaces, because the exact gauge-invariance is recovered only in the whole space. An optimal gauge can be defined as the one predicting the most accurate observables for the same number of atomic levels and modes. Different metrics quantifying the gauge performance can be introduced depending on the observable of interest. In this work we demonstrate that the optimal choice is generally mode-dependent, i.e., a different gauge is needed for each cavity mode. While the choice of gauge becomes more important for increasing light-matter interaction, we also show that the optimal gauge does not correspond to the situation where the entanglement between light and matter is the smallest.

\end{abstract}
\date{\today}
\maketitle

\section{Introduction}
Cavity \cite{Haroche2006} and circuit quantum electrodynamics \cite{RMP_circuitQED} (QED) are branches of quantum physics that have attracted a great deal of interest for a variety of fundamental quantum phenomena and for applications in quantum information, thanks to the manipulation of atoms by quantized electromagnetic fields and viceversa. In many different platforms, it is nowadays possible to enhance vacuum fields by spatial confinement and achieve non-perturbative light-matter interactions between atoms and resonators \cite{Forn_2019,kockum2019ultrastrong,garcia2021manipulating}. For practical calculations, one in general is obliged to truncate the Hilbert space by reducing the number of atomic levels, the number of modes or the number of photons in each mode. Gauge-invariance is an important property that holds in the global Hilbert space, but it is lost when working in a subspace \cite{Bassani_1977,Cohen_Tannoudji_1997,Nazir2020b}. It is important to note that some physical observables strongly depend on the gauge: in the historic example of the 1s-2s two-photon absorption for the hydrogen atom, the Coulomb gauge gives a zero effect in a two-level approximation and, by increasing the number of levels,  converges much slower than the dipole gauge to the exact result \cite{Bassani_1977}. 

Ultra-strong light-matter interactions exacerbate these gauge subtleties: several works in the literature have been devoted to the so-called gauge ambiguities \cite{De_Bernardis_2018b,roth2019optimal,dmytruk2021gauge,Di_Stefano_2019,Nazir2020b}, especially in the context of the quantum Rabi model, where the atom is approximated by a two-level system and the cavity field has a single mode.  For practical calculations, one can introduce the concept of optimal gauge, as done in a recent circuit QED work \cite{roth2019optimal}. In general the metric will depend on the observables of interest, like for example the energy spectrum.  In Ref. \cite{roth2019optimal}, it was found that an optimal gauge is in general a mixed gauge in-between the dipole and Coulomb gauge and was taken to be the same for all modes.  Another work \cite{Ashida_2021} introduced a transformation that produces light-matter decoupling for large couplings with the goal to systematically derive low-energy effective models. Recent multi-mode circuit QED works considered  for convenience a gauge of one kind (flux gauge) for a set of low-frequency modes and a second gauge (charge gauge) for a set of high-frequency modes \cite{Nitish_paper1,Nitish_paper2}. The state-of-the-art points to an emergent fundamental interest for optimal gauges regarding their efficiency, their properties and their relation with light-matter entanglement.

The goal of this article is to investigate in a rigorous way some key properties of optimal gauges by comparing the predictions of truncated multi-mode cavity QED models to the corresponding exact results. The manuscript is organized as follows. In Sec. \ref{framework}, we introduce the theoretical framework, focusing on the Hamiltonian model describing an atom coupled to a multi-mode resonator. In Sec. \ref{results}, we presents results concerning the optimal atomic basis truncation for a given gauge (Sec. \ref{truncation}), the optimal gauge for single-mode (Sec. \ref{single}) and multi-mode cavities (Sec. \ref{multi}). Moreover, we study the relation between light-matter entanglement and the optimal gauges in Sec. \ref{entanglement_section}. Conclusions and perspectives are drawn in Sec. \ref{conclusions}.

\section{Theoretical framework}
\label{framework}
 In order to be able to compare exact results to truncated models, we will consider the Hamiltonian model describing an atom with a single degree of freedom \cite{De_Bernardis_2018b,roth2019optimal}, as represented by the Hamiltonian 
\begin{equation}
\hat{\mathcal{H}}_0 \ = \  
\frac{ \hat{p}^2 }{2 m} + V(\hat{x})\, ,
\end{equation}
where  $V(\hat{x})$ is the potential energy depending on the coordinate $\hat{x}$ and $\frac{ \hat{p}^2 }{2 m}$ the kinetic energy operator depending on the conjugate momentum operator $\hat{p}$ and mass $m$.  In the Coulomb gauge, the Hamiltonian of the atom in the presence of a quantum electromagnetic field described by the vector potential operator $\hat{\mathbf{A}}$ reads
\begin{equation}
\hat{\mathcal{H}}_C \ = \  
\frac{1}{2 m} ( \hat{p} + q \hat{A} )^2 + V(\hat{x}) +  \hat{\mathcal{H}}_{p}\, ,
\end{equation}
where $\hat{A}$ is the component of the vector potential along the considered atom dimension and $q$ is the charge. The operator $\hat{\mathcal{H}}_{p}=\sum_k \hbar \omega_k  \hat{b}_k^{\dagger}\hat{b}_k $ is the bare Hamiltonian of the photon modes with frequencies $\omega_k$ and bosonic creation (destruction) operators  $\hat{b}_k^{\dagger} (\hat{b}_k)$.
Assuming the spatial size of the atom is much smaller than that of the electromagnetic modes, the spatial dependence of the vector potential can be ignored. In this limit, we can rewrite $\hat{A} = \sum A_k  (\hat{b}_k + \hat{b}_k^{\dagger})$, i.e., not depending on the operator $\hat{x}$. Under this approximation, the Coulomb gauge Hamiltonian reads:
\begin{equation}
\hat{\mathcal{H}}_C  = \hat{\mathcal{H}}_0 + \frac{\hat{p}}{ m}  \sum_k  q  A_k ( \hat{b}_k +   \hat{b}_k^{\dagger}   )   +   \frac{q^2}{2m} \left[   \sum_k   A_k  ( \hat{b}_k +   \hat{b}_k^{\dagger}   ) \right]^2 +  \hat{\mathcal{H}}_{p}  \, .
\end{equation}
Another common choice is the dipole gauge, where the Hamiltonian is given by:
\begin{equation}
\hat{\mathcal{H}}_D = 
 \hat{\mathcal{H}}_0 + \sum_k \frac{ \omega_k q^2 A_k^2 }{\hbar} \hat{x}^2 
-i \hat{x} \sum_k  \omega_k q A_k  ( \hat{b}_k^{\dagger} - \hat{b}_k )    + \hat{\mathcal{H}}_{p} .
\end{equation}
In the dipole gauge the coupling between the atom and the cavity occurs via the position operator $\hat{x}$, while in the Coulomb gauge it is via the momentum operator $\hat{p}$.
As the two Hamiltonians describe exactly the same physical system, the two are related by an unitary transformation, namely $\hat{\mathcal{H}}_D = \hat{U}  \hat{\mathcal{H}}_C \hat{U}^\dagger  $ where 
${\hat{U} = \exp{ i q \hat{x} \hat{A} /\hbar }}$ 
is the Power-Zienau-Woolley (PZW) transformation \footnote{In principle, the gauge should be fixed before the electromagnetic field is quantized. However, in non-relativistic QED one can quantize the theory while keeping the gauge arbitrary \cite{Cohen_Tannoudji_1997}}.
Other than the Coulomb and dipole gauges, there is an infinite number of possible choices. A subset of possible gauge transformations can be generated by the unitary operator
${\hat{U}_\eta = \exp{ i \eta q \hat{x} \hat{A} /\hbar }}$ \cite{stokes2019gauge} giving the corresponding Hamiltonian $ \hat{\mathcal{H}}_\eta = \hat{U}_\eta \hat{\mathcal{H}}_C \hat{U}_\eta^{\dagger} $. The parameter $ \eta$ interpolates continuously between the Coulomb ($\eta=0$) and dipole ($\eta=1$) gauges. 
The resulting Hamiltonian can be written as
$\hat{\mathcal{H}}_\eta =  \hat{\mathcal{H}}^a_\eta + \hat{\mathcal{H}}^{int}_\eta + \hat{\mathcal{H}}^p_\eta $ with 
\begin{equation}
\hat{\mathcal{H}}^a_\eta =  
\frac{ \hat{p}^2 }{2 m} + V(\hat{x}) + \eta^2 \sum_k \frac{ \omega_k q^2 A_k^2 }{\hbar} \hat{x}^2  \, ,
\end{equation}
\begin{equation}
 \hat{\mathcal{H}}^{int}_\eta  =
(1-\eta) \frac{\hat{p}}{ m}  \sum_k  q  A_k ( \hat{b}_k +   \hat{b}_k^{\dagger}   ) 
-i \eta \hat{x} \sum_k  \omega_k q A_k  ( \hat{b}_k^{\dagger} - \hat{b}_k )   \, ,
\end{equation}
\begin{equation}
\hat{\mathcal{H}}^p_\eta  = 
  \frac{1}{2m} \left[  (1-\eta) \sum_k  q A_k  ( \hat{b}_k +   \hat{b}_k^{\dagger}   ) \right]^2 +  \sum_k \hbar \omega_k  \hat{b}_k^{\dagger}\hat{b}_k \, .
\end{equation}
While different gauges give rise to the same physics, each part of the Hamiltonian is not gauge invariant. Indeed, the atomic part, the photonic part and the interaction part are all different in different gauges.
The purely photonic part is quadratic in the creation and annihilation operators and thus can be diagonalized by a Bogoliubov transformation.

The derivation of effective low-energy models is a cornerstone of modern condensed matter physics \cite{Girvin_2019}. Formally, effective models are obtained by projecting the full Hamiltonian $\hat{\mathcal{H}}$ to a lower dimensional subspace. The reduced Hamiltonian is obtained via the projection $ \hat{\mathcal{H}}_r  = \hat{\mathcal{P}} \hat{\mathcal{H}}  \hat{\mathcal{P}}$ where $ \hat{\mathcal{P}} = \sum_j | \Psi_j \rangle \langle \Psi_j | $ with $ | \Psi_j \rangle $ being a (finite or infinite) set of orthonormal states.
In general, the truncation of the Hilbert space produces a breakdown of the gauge invariance \cite{Bassani_1977,De_Bernardis_2018b,roth2019optimal,stokes2019gauge}.

{\it Gauge-dependent light-matter separability.---}
Suppose that $\hat{\mathcal{H}}$ and $\hat{\mathcal{H}}'$  correspond to the same system in two different gauges and that they are linked  by a unitary transformation $\hat{U}$ such that $\hat{\mathcal{H}} = \hat{U}^{\dagger} \hat{\mathcal{H}}'  \hat{U}$. Note that $ \hat{\mathcal{P}}  \hat{\mathcal{H}}' \hat{\mathcal{P}} $ and $   \hat{\mathcal{P}} \hat{ \mathcal{H}} \hat{\mathcal{P}} $ are {\it not} connected by a unitary transformation and do not produce the same energy spectrum. This conclusion merely reflects the fact that $ \hat{\mathcal{P}}  $ is not gauge invariant. Since the transformation between the different gauges is done by a unitary transformation, if one wants to consider an equivalent reduced model, also the projection operator needs to be accordingly transformed as $ \hat{\mathcal{P}}' = \hat{U}^\dagger  \hat{\mathcal{P}} \hat{U}$. The reduced models $\hat{\mathcal{P}}' \hat{\mathcal{H}}'  \hat{\mathcal{P}}' $ and $ \hat{\mathcal{P}}   \hat{\mathcal{H}}  \hat{\mathcal{P}} $ are physically equivalent and with the same spectrum. However, if one wants to truncate only the atomic part of the Hilbert space, as typically done,  one has to consider the projection $ \hat{\mathcal{P}}= \hat{P} \otimes \mathds{1}_p $, where $ \hat{P} = \sum_{j=1}^M | j \rangle \langle j | $ is a projector over a finite set of atomic levels, and $ \mathds{1}_p$ is the identity operator for the photonic Hilbert space.
It is important to note that $\hat{\mathcal{P}}'$ generally cannot be written in the factorized form $ \hat{P}' \otimes \mathds{1}_p $ as it was assumed for $\hat{\mathcal{P}}$. Namely, the projection to a set of separable states in a given gauge in general corresponds to a projection to a set of entangled states in any other gauge. To avoid this complication, one can consider only projections that are separable, with respect to the chosen gauge. However, the reduced models obtained under this restriction are not gauge invariant, thus implying the existence of an optimal gauge in terms of accuracy.

Few recent works \cite{Di_Stefano_2019,dmytruk2021gauge,taylor2020} have claimed that reduced models can be made gauge invariant by introducing a projected gauge-fixing transformation. The essence of this approach is to transform the Hamiltonian using $  \hat{\mathcal{P}} \hat{ U} \hat{\mathcal{P}}  $ \cite{Di_Stefano_2019} or  $ \hat{ U} ( \hat{\mathcal{P}} \hat{ x} \hat{\mathcal{P}})  $ \cite{taylor2020}. The latter transformation is unitary, and is restricted to the projected subspace.
While the projected Hamiltonian is invariant under this transformation, it still depends on the initial choice of gauge, where the truncation operation have taken place.

\section{Results and discussion}
\label{results}
\subsection{Optimal atomic basis for a given gauge}
\label{truncation}

Before we investigate the optimal gauge, we start with the problem of how to perform the truncation of the atomic Hilbert space. As shown in the previous section, the interaction with the cavity leads to a renormalization of the atomic potential, which is given by:
\begin{equation}
V^{\mathrm{(eff)}}_{\eta}(\hat{x}) \ = \ V(\hat{x}) + \eta^2 \sum_k \frac{ \omega_k q^2 A_k^2 }{\hbar} \hat{x}^2 . \label{veff}
\end{equation}
The additional gauge-dependent term (i.e., depending on $\eta$) in the renormalized potential $V^{\mathrm{(eff)}}_{\eta}(\hat{x})$ for strong interactions or many modes can distort significantly the bare potential $V(\hat{x})$.
Since the eigenstates of $\hat{\mathcal{H}}^a_\eta$ are  different from those of the bare atomic Hamiltonian $\hat{\mathcal{H}}_0$ without the cavity, one can introduce  a gauge-dependent projection \cite{stokes2019gauge} operator $ \hat{P}_{\eta} $ that truncates the Hilbert space to the lowest energy levels of $ \hat{\mathcal{H}}^a_\eta $. As the truncated atomic levels now depend on the gauge choice, one might naively expect to obtain a more accurate model by considering the eigenstates of the atomic Hamiltonian with the renormalized potential. Here we show that, surprisingly, this is not the case.  To demonstrate that, we consider the celebrated quantum Rabi model, which in recent works has been extensively investigated in the context of gauge invariance \cite{De_Bernardis_2018b,Di_Stefano_2019,roth2019optimal,Nazir2020b,stokes2019gauge}. The quantum Rabi model is obtained by truncating the atomic Hilbert space to the two lowest-energy levels. To address a concrete example, let us study the case of a double well potential given by
\begin{equation}
V(\hat{x})=C\hat{x}^4 - B \hat{x}^2 \, ,
\label{double_well}
\end{equation}
with $B,C > 0$. A simple scaling analysis shows that the spectral anharmonicity of the bare Hamiltonian $\hat{\mathcal{H}}_0$ depends on a single dimensionless parameter $\gamma = \frac{m B^3}{ \hbar^2 C^2} $. Increasing the value of $\gamma$, the anharmonicity of the spectrum is enhanced. In particular, the two lowest energy levels can be well separated from higher excited levels by increasing $\gamma$ enough.

We denote by $| \epsilon_0^{(\eta)} \rangle $ and $| \epsilon_1^{(\eta)} \rangle $ the eigenstates of the renormalized atomic Hamiltonian $\hat{\mathcal{H}}^a_\eta$ corresponding to the two lowest-energy eigenvalues $\epsilon_0^{(\eta)}$ and $\epsilon_1^{(\eta)}$.
The projection operator to the corresponding two-dimensional subspace is given by $ \hat{\mathcal{P}}_\eta= \hat{P}_\eta \otimes \mathds{1}_p $ with $\hat{P}_\eta=| \epsilon_0^{(\eta)} \rangle \langle \epsilon_0^{(\eta)} | + | \epsilon_1^{(\eta)} \rangle \langle \epsilon_1^{(\eta)} | $.
The corresponding quantum Rabi model reads:
\begin{eqnarray}
%
\hat{\mathcal{P}}_{\eta} \hat{\mathcal{H}}_{\eta} \hat{\mathcal{P}}_{\eta}  =  
  & -&\frac{\Delta^{(\eta)}}{2}  \hat{\sigma}_z + (1-\eta) \sum_k g_C^{\eta,k} \hat{\sigma}_y  ( \hat{b}_k +  \hat{b}_k^{\dagger} ) \nonumber \\ &-& i \eta \sum_k g_D^{\eta,k} \hat{\sigma}_x  ( \hat{b}_k^{\dagger} - \hat{b}_k )  + \hat{\mathcal{H}}^p_\eta    \, , 
\end{eqnarray}
where $ \Delta^{(\eta)}  = \epsilon_1^{(\eta)} - \epsilon_0^{(\eta)}  $ , $g_C^{\eta,k} = i q A_k \langle \epsilon_0^{(\eta)} | \hat{p}  | \epsilon_1^{(\eta)} \rangle /m $ and $g_D^{\eta,k} = \omega_k q A_k \langle \epsilon_0^{(\eta)} | \hat{x}  | \epsilon_1^{(\eta)} \rangle $. 

Alternatively, we can use the projector $ \hat{\mathcal{P}}_0 = \hat{P}_0  \otimes \mathds{1}_p $, with $\hat{P}_0 =| \epsilon_0  \rangle \langle \epsilon_0  | + | \epsilon_1  \rangle \langle \epsilon_1  | $, where $| \epsilon_0 \rangle $ and $| \epsilon_1 \rangle $ are the lowest energy levels of the bare atomic Hamiltonian $\hat{\mathcal{H}}_0$, namely $\epsilon_0 = \epsilon_0^{(\eta = 0)}$ and $\epsilon_1 = \epsilon_1^{(\eta = 0)}$.
In this case we obtain:
\begin{eqnarray}
%
\hat{\mathcal{P}}_0 \hat{\mathcal{H}}_\eta \hat{\mathcal{P}}_0 \ = \
  &-& \left(\frac{\Delta}{2}  +  \sum_k \eta^2 \delta_k \right)  \hat{\sigma}_z  \nonumber \\ &+&  \sum_k (1-\eta) g_C^{k} \hat{\sigma}_y  ( \hat{b}_k +  \hat{b}_k^{\dagger} ) \nonumber  \\ &-& i \sum_k \eta  g_D^{k} \hat{\sigma}_x  ( \hat{b}_k^{\dagger} - \hat{b}_k )  + \hat{\mathcal{H}}^p_\eta    \, . \label{P0HP0}
\end{eqnarray}
Here $ \Delta = \epsilon_1 - \epsilon_0 $ , $g_C^k = i q A_k \langle \epsilon_0 | \hat{p}  | \epsilon_1 \rangle /m $ and $g_D^k = \omega_k q A_k \langle \epsilon_0 | \hat{x}  | \epsilon_1 \rangle $.
Note that the additional corrections 
\begin{eqnarray}
 \delta_k = \frac{ \omega_k q^2 A_k^2 }{\hbar}  ( \langle \epsilon_1  | \hat{x}^2  | \epsilon_1  \rangle  - \langle \epsilon_0 | \hat{x}^2  | \epsilon_0 \rangle ) /2
\end{eqnarray}
results from the projection of the renormalization term proportional $\hat{x}^2$.
\begin{figure}[!ht]
\includegraphics[width=1\hsize]{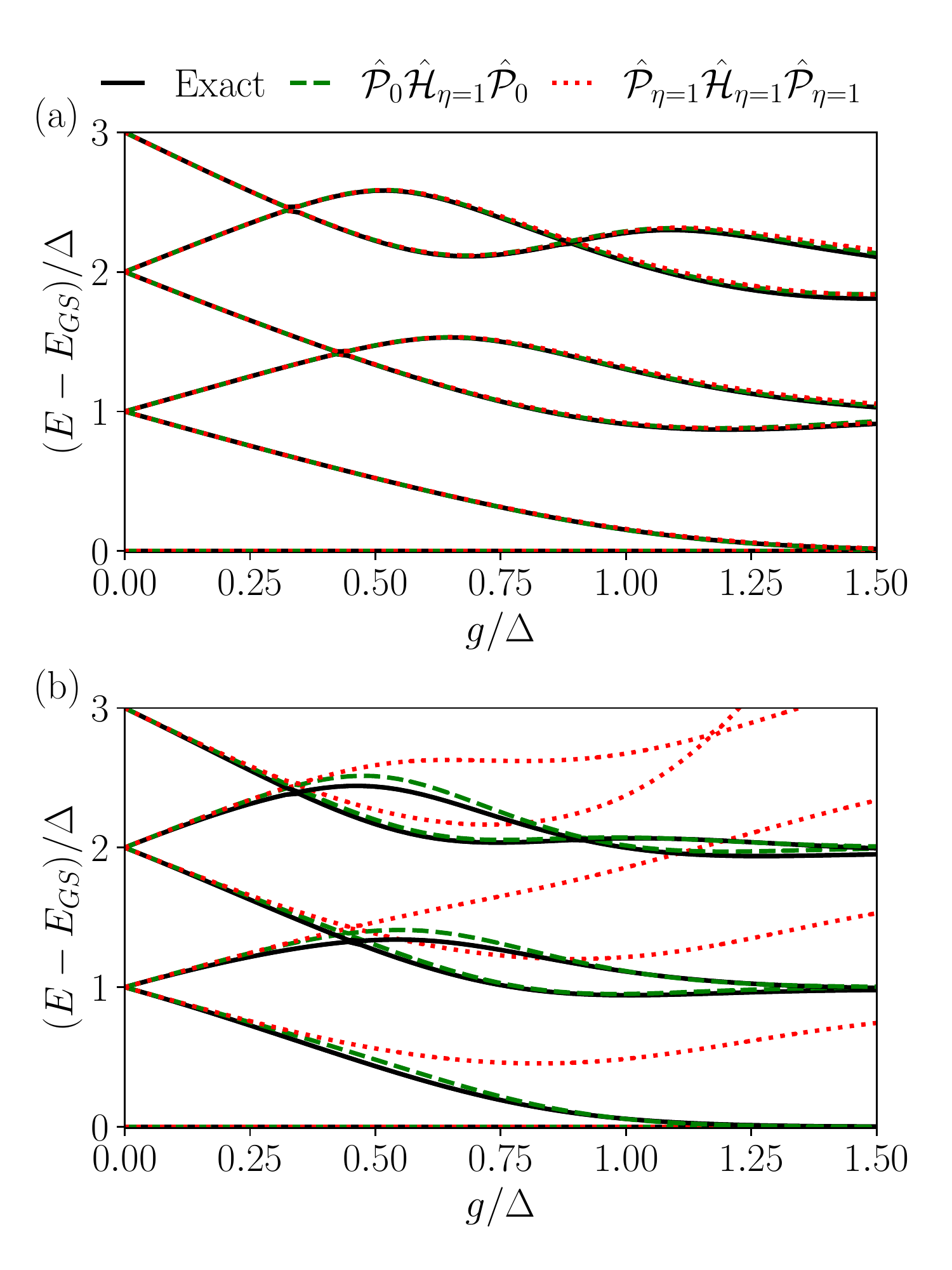}
\caption{Comparison of the light-matter energy eigenvalues (measured with respect to the ground state energy) in the dipole gauge ($\eta = 1$) for the full Hamiltonian (exact results, black lines), for the model with the truncation in the bare atomic basis ($\hat{\mathcal{P}}_0 \hat{\mathcal{H}}_\eta \hat{\mathcal{P}}_0$, green dashed lines) and for that truncated in the renormalized atomic basis ($\hat{\mathcal{P}}_{\eta} \hat{\mathcal{H}}_\eta \hat{\mathcal{P}}_{\eta}$, red dotted lines). (a) single-mode cavity with $\hbar \omega_1=\Delta = \epsilon_1 - \epsilon_0 $ (cavity mode resonant to the energy transition between the bare atomic ground state and first excited level). (b) two-mode cavity with $\hbar \omega_1=\Delta$ and $\hbar \omega_2 = 20\Delta $.  The bare atomic potential is given by Eq. (\ref{double_well}) with anharmonicity parameter $\gamma= 64 $,
giving $(\epsilon_2-\epsilon_0)/(\epsilon_1-\epsilon_0) \simeq 26$, where $\epsilon_0$, $\epsilon_1$ and $\epsilon_2$ are the three lowest energy eigenvalues of the bare atomic Hamiltonian.  \label{fg1} }
\end{figure}

Now, let us compare the spectra of $\hat{\mathcal{P}}_\eta \hat{\mathcal{H}}_\eta \hat{\mathcal{P}}_\eta $ and $\hat{\mathcal{P}}_0 \hat{\mathcal{H}}_\eta \hat{\mathcal{P}}_0$, which are the quantum Rabi models obtained by using  the two lowest energy eigenstates of respectively the renormalized and bare atomic Hamiltonian. In the following, we  will assume for simplicity $A_k=A_1$ and consider only up to three modes in order to have exact results for the full cavity QED model. Note that depending on the spatial position of the atom with respect to the spatial mode profiles, it is possible in cavity systems to tailor the relative weight of the mode vacuum fields. With this assumption for the mode vacuum field amplitudes,  we can use a single parameter
\begin{eqnarray}
g = \frac{qA_1 | \langle \epsilon_0 | p  | \epsilon_1 \rangle | }{m} \, 
\end{eqnarray}
to characterise the interaction strength in single or multi-mode cavities. Of course, our theory can be applied to any arbitrary set of vacuum field amplitudes $A_k$.

{\it Exact calculations.---}
To benchmark the behavior of the different gauges and truncations, we have calculated the numerically exact energy eigenvalues with the full Hamiltonian by discretizing the values of the spatial coordinate $x$ and by introducing a cutoff for the number of photons in each mode. The convergence in the continuum limit has been carefully verified by decreasing the spatial grid step and by increasing the photon number cutoff. We have also carefully verified that we get the same energy spectrum for every value of $\eta$, that is for every gauge, as it must be.

{\it Bare versus renormalized atomic basis truncation.---} In Fig.\ref{fg1} we compare the lowest energy eigenvalues of $\hat{\mathcal{P}}_\eta \hat{\mathcal{H}}_\eta \hat{\mathcal{P}}_\eta $ and $\hat{\mathcal{P}}_0 \hat{\mathcal{H}}_\eta \hat{\mathcal{P}}_0  $ against the exact eigenvalues for the full Hamiltonian.  Here we fix $\eta=1$, which corresponds to the dipole gauge, where the renormalization term for the atomic potential is the largest. Note that instead for $\eta = 0$ (Coulomb gauge), $V^{(eff)}_{\eta = 0}(x) = V(x)$. For $\eta  =1$, we first consider a single-mode cavity (top panel), where the mode is resonant with the atomic transition. In this case both reduced models fit well the exact spectrum, and the difference between the two can barely be resolved, even for strong light-matter interaction energy $g$ (compared to the atomic transition energy $\Delta$).  However, when we add a second mode of high frequency (bottom panel) we clearly see that $\hat{\mathcal{P}}_0 \hat{\mathcal{H}}_\eta \hat{\mathcal{P}}_0 $ provides a much better agreement compared to $\hat{\mathcal{P}}_\eta \hat{\mathcal{H}}_\eta \hat{\mathcal{P}}_\eta $. Namely, projecting to the bare atomic level basis $ | \epsilon_i  \rangle  $ provides a much better agreement with the exact continuum model, while the basis $ | \epsilon_i^{(\eta)} \rangle  $ provides a truncation basis that is very inaccurate. 

{\it Shortcomings of renormalized atomic basis.---}
The fact that the truncation in the renormalized atomic basis introduces a significant error is not so surprising once we inspect Eq. (\ref{veff}) and see that all modes can contribute to the renormalization of the atomic potential, even when the photon energy $\hbar \omega_k$ is much larger than the light-matter interaction energy $g$ and the atomic transition energy $\Delta$.
Another additional argument explaining the significant error associated to the truncation on the renormalized atom basis ($\hat{\mathcal{P}}_\eta \hat{\mathcal{H}}_\eta \hat{\mathcal{P}}_\eta $) can be attributed to the ``arbitrariness" of the renormalized potential in Eq. (\ref{veff}), as we explain below. Suppose the atomic "trapping`` potential is shifted by a constant distance $d$, such that $V(\hat{x}) \rightarrow V(\hat{x}-d)$. The bare atomic spectrum is not changed by this translation, while the new eigenstates are related to the old ones by a simple translation. The same applies to the full Hamiltonian in the presence of the cavity field, assuming the cavity mode amplitude remain the same in the shifted position.
It would be natural to require that the projected Hamiltonian would also satisfy this trivial symmetry. However, if we project the Hamiltonian using the $ | \epsilon_i^{(\eta)} \rangle $ basis, the results would depend on the constant $d$. This is because the second term in the effective potential remains unchanged, so that $V^{(\mathrm{eff})}_{\eta}(\hat{x}) \not\rightarrow V^{(\mathrm{eff})}_{\eta}(\hat{x} - d)$. Hence the projected model $\hat{\mathcal{P}}_\eta \hat{\mathcal{H}}_\eta \hat{\mathcal{P}}_\eta $ would depend on the arbitrary constant $d$, unlike $\hat{\mathcal{P}}_0 \hat{\mathcal{H}}_\eta \hat{\mathcal{P}}_0 $.  Furthermore, the effective potential can be also modified by a unitary transformation that acts only on the photonic part. For instance, the transformation $ U = \mathrm{e}^{i s ( \hat{b}_1+  \hat{b}_1^{\dagger}) }$ would add the term $ \eta s \omega_1 q A_1 \hat{x} $  to $V^{(\mathrm{eff})}_{\eta}(\hat{x}) $, and thus change the levels $ | \epsilon_i^{(\eta)} \rangle  $ and the spectrum of the projected model $\hat{\mathcal{P}}_\eta \hat{\mathcal{H}}_\eta \hat{\mathcal{P}}_\eta $. 
To conclude, although $V^{(\mathrm{eff})}_{\eta}(\hat{x}) $ does not involve photonic operators, it should not be interpreted as a pure atomic potential and the truncation is its basis of energy eigenstates can lead to strongly inaccurate results.

Note that previous works in the literature have focused on a single-mode cavity and on the resonant case $\hbar \omega_1 = \Delta$ with a highly anharmonic double well potential for the atom \cite{De_Bernardis_2018b,Di_Stefano_2019}: in this configuration the truncation in the renormalized atom basis turns out to be a good approximation even for very strong interaction strengths.
For the double well potential, the effective renormalization of the potential is negligible when $  \sum_k \omega_k q^2 A_k^2 /\hbar   \ll B $. For the anharmonicity of $\gamma=64$ considered in the figures discussed above, the two sides of this inequality are equal when $g/\Delta \sim 8$. Therefore, in Fig.\ref{fg1}(a) we have $\hat{\mathcal{P}}_\eta \approx \hat{\mathcal{P}}_0  $. Indeed, the difference between the two projected models can be barely resolved.

\begin{figure}[t!]
\includegraphics[width=1\hsize]{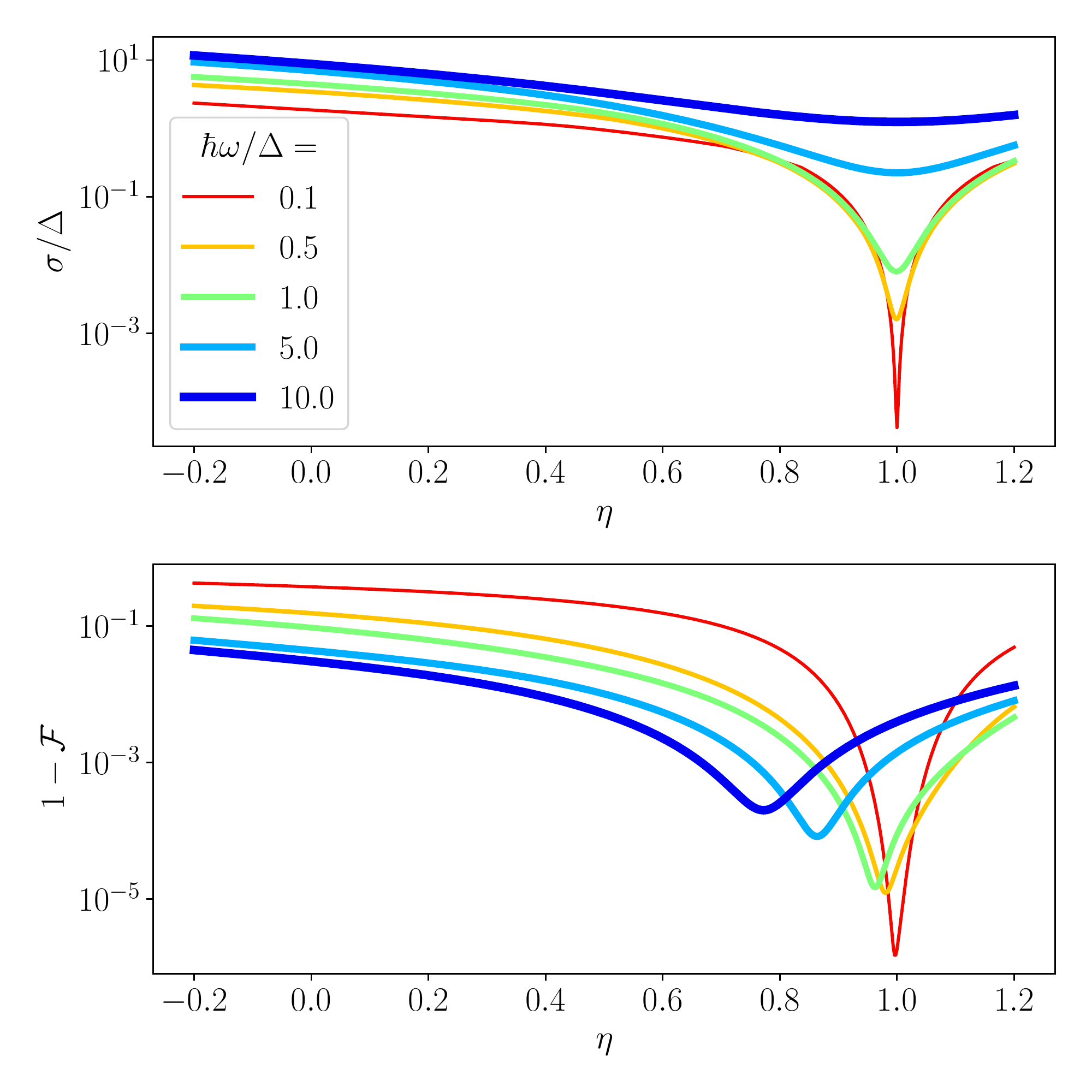}
\caption{Optimal gauge for a single-mode cavity. Top panel: the energy spectrum deviation $\sigma$, defined in Eq. (\ref{sigma}), between the exact spectrum and that predicted by the truncated model, as a function of the gauge parameter $\eta$. Bottom panel: the ground state infidelity $1 - {\mathcal F}$ between the truncated and exact models, as a function of $\eta$. The different curves correspond to different cavity mode frequencies, as indicated in the legend. The interaction strength is $g=0.8 \Delta$ in all plots (see definition in the text). Other parameter: $M = 7$ (we  consider the first $7$ excitation energies of the cavity QED Hamiltonian).} \label{fgMetrics}
\end{figure}

\begin{figure*}[!ht]
\includegraphics[width=1\hsize]{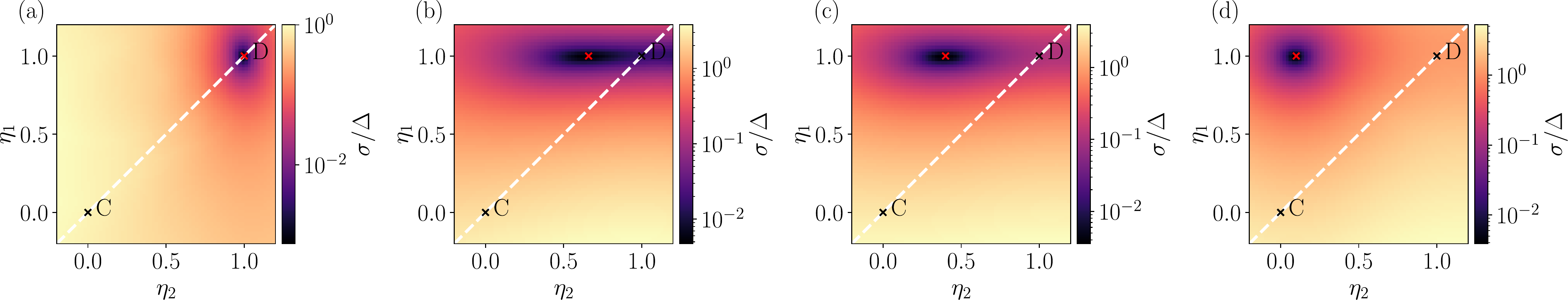}
\caption{Optimal gauge for two-mode cavity systems. The energy spectrum deviation $\sigma$ is plotted as a function of the gauge parameters $\eta_1$ and $\eta_2$ corresponding to the two modes. Panels (a-d) are for a two-mode cavity with frequencies $\hbar \omega_k / \Delta = (1,0.5) $(a), $ (1,10) $(b), $ (1,30) $(c) and $ (1,200) $(d). The interaction strength is $g = 0.6 \Delta $ in all panels. The dashed lines indicate the uniform gauge condition where $\eta_1=\eta_2$. The standard Coulomb (C) and Dipole (D) gauges are indicated in the figure. The optimal gauge is indicated by a red cross marker.
\label{multimode}
}
\end{figure*}

\begin{figure}
\includegraphics[width=1\hsize]{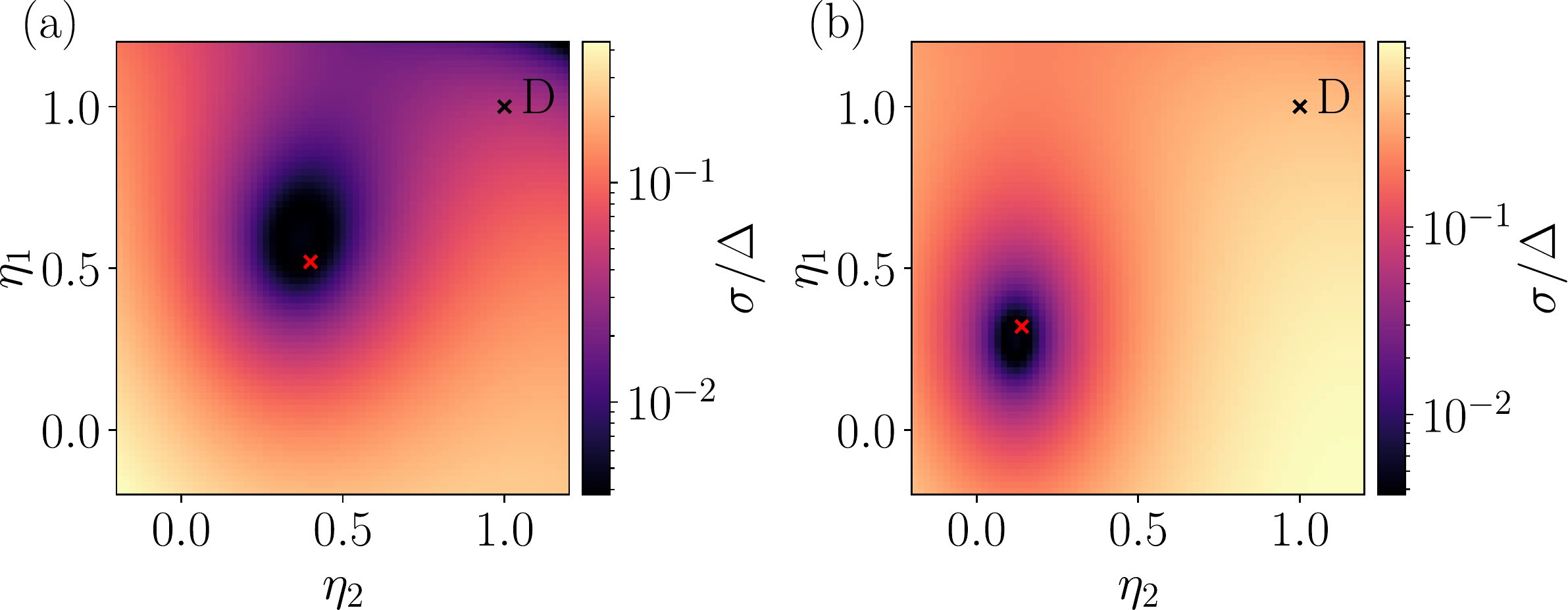}
\caption{Optimal gauge for three-mode cavity systems. The energy spectrum deviation $\sigma$ is plotted as a function of the gauge parameters $\eta_1$ and $\eta_2$ corresponding to the first two modes. Panels (a-b) are for a cavity with frequencies $\hbar \omega_k / \Delta = (10,30,1) $(a) and $ (50,150,1) $(b). The interaction strength is $g = 0.6 \Delta $ in both panels.
\label{threemode}
}
\end{figure}

\subsection{Optimal gauge for a single-mode cavity}
\label{single}
Having clarified that the reduced model is gauge dependent and how to choose the basis for truncation (the bare one), we now address the problem of finding the optimal gauge. Namely, we search for the gauge where $ \hat{\mathcal{P}}_0 \hat{\mathcal{H}}_\eta \hat{\mathcal{P}}_0$ best represents the low energy physics of $\hat{\mathcal{H}}_\eta  $.
It should be noted that the comparison of different gauges depends on the observable of interest, and therefore different metrics can be used. The difference between the low-energy spectrum of the full and the reduced models can be quantified for example by the standard deviation 
\begin{equation}
\sigma = \sqrt{ \sum_{i=1}^M (E_i - E_i^{(\eta)})^2 / M } \, ,
\label{sigma}
\end{equation}
which involves the first $M$ excitation energies $E_i$ ($E_i^{(\eta)}$) with respect to the ground state of the light-matter system for the full (truncated) cavity QED model \footnote{The results presented here are not very sensitive to the choice of $M$.}.
In order to quantify the accuracy of the ground state wave-function, we have also considered the ground-state fidelity 
\begin{equation}
\mathcal{F} = | \langle \psi_0^{(\eta)} | \Psi_0^{(\eta)} \rangle |^2 \, ,
\end{equation}
where $ |  \psi_0^{(\eta)} \rangle $ and $ | \Psi_0^{(\eta)} \rangle  $ are the ground states of the truncated model and the full Hamiltonian both in the same $\eta$-gauge.

In Fig. \ref{fgMetrics}, we plot the spectral deviation $\sigma$ and the ground-state fidelity $\mathcal{F}$ for a single-mode cavity for different mode frequencies as a function of the gauge parameter $\eta$. When comparing the energy spectrum (Fig. \ref{fgMetrics}, top panel), we find that the dipole gauge ($\eta = 1$) always produces the best results, in accordance with previous studies \cite{De_Bernardis_2018b}.

The behavior of the ground state fidelity, plotted in the bottom panel of Fig. \ref{fgMetrics}, is starkly different from the spectral deviation $\sigma$. Indeed, even with only one mode, the optimal gauge best approximating the ground state is not the dipole gauge. In particular, by increasing the cavity mode frequency the optimal $\eta$ decreases and the overall accuracy decreases.

\subsection{Mode-dependent optimal gauge}
\label{multi}
For a single-mode cavity, we have seen earlier that the reduced model in the dipole gauge provides the most accurate spectrum, in agreement with previous studies \cite{De_Bernardis_2018b}. However, when we have more than one mode this is no longer the case.
A two-mode quantum Rabi model was studied in Ref. \cite{roth2019optimal}, where it was shown that the optimal gauge is neither the Coulomb gauge nor the dipole one, but rather some intermediate gauge such that $0<\eta<1$. However, in \cite{roth2019optimal} the same gauge was assumed for each mode. The case of a Josephson atom coupled to a manifold of modes in a transmission line resonator was recently explored in \cite{Nitish_paper1}, where one gauge was used for a set of low frequency modes and one another gauge was taken for a set of high-frequency modes. Yet, given the complexity of the system, it was not investigated which gauge was optimal.

Here, we address this problem in the considered framework by replacing $\eta$ by the set $ \{ \eta_i \} $, thus allowing a different gauge for every cavity mode. Let us now introduce the mode-dependent transformation:
\begin{eqnarray}
\hat{U}_{ \{  \eta \}   } \ = \  \prod_k \exp{ i \eta_k q \hat{x} A_k  (\hat{b}_k + \hat{b}_k^{\dagger}) /\hbar } \,.
\end{eqnarray}
The transformed Hamiltonian $\hat{\mathcal{H}}_{\{ \eta \} } = \hat{U}_{\{ \eta \} } \hat{\mathcal{H}}_C \hat{U}_{\{ \eta \} }^{\dagger} $ can be written as $\hat{\mathcal{H}}_{\{ \eta \} } =  \hat{\mathcal{H}}^a_{\{ \eta \} } + \hat{\mathcal{H}}^{int}_{\{ \eta \} } + \hat{\mathcal{H}}^p_{\{ \eta \} } $ with 
\begin{equation}
\hat{\mathcal{H}}^a_{\{ \eta \} } =  
\frac{ \hat{p}^2 }{2 m} + V(\hat{x}) +  \sum_k \eta_k^2 \frac{ \omega_k q^2 A_k^2 }{\hbar} \hat{x}^2  \, ,
\end{equation}
\begin{equation}
 \hat{\mathcal{H}}^{int}_{\{ \eta \} }  =
 \frac{\hat{p}}{ m}  \sum_k (1-\eta_k) q  A_k ( \hat{b}_k +   \hat{b}_k^{\dagger}   ) 
-i  \hat{x} \sum_k \eta_k \omega_k q A_k  ( \hat{b}_k^{\dagger} - \hat{b}_k )   \, ,
\end{equation}
\begin{equation}
\hat{\mathcal{H}}^p_{\{ \eta \} }  = 
  \frac{1}{2m} \left[  \sum_k  (1-\eta_k) q A_k  ( \hat{b}_k +   \hat{b}_k^{\dagger}   ) \right]^2 +  \sum_k \hbar \omega_k  \hat{b}_k^{\dagger}\hat{b}_k \, .
\end{equation}
Once again, our goal is to compare exact results to the prediction of gauge-dependent truncated models. For the mode-dependent gauge, $\hat{\mathcal{P}}_0 \hat{\mathcal{H}}_{\{ \eta \} } \hat{\mathcal{P}}_0$ is given by Eq. (\ref{P0HP0}) with $\eta \rightarrow \eta_k$ and $ \hat{\mathcal{H}}^p_{ \eta  } \rightarrow \hat{\mathcal{H}}^p_{\{ \eta \} }$.
Similarly, we define the spectrum deviation $\sigma$ as in Eq. (\ref{sigma}), where here $E_i^{ ( \eta) } \rightarrow E_i^{ \{ \eta \} }  $.

In Fig. \ref{multimode} we plot the spectrum deviation $\sigma$ as a function of the gauge parameters $ \{\eta_i \}$ for the case of a two-mode cavity. In all panels we keep one mode frequency resonant with the first atomic transition. As shown in panel (a), when the second mode has a low frequency with the respect to the atomic transition, the dipole gauge for both modes $(\eta_1=\eta_2=1)$ produces the most accurate model, as in the single-mode case. However, as reported in panels (b-d), when the second mode frequency is high with respect to the atomic transition frequency $\Delta$, this is no longer the case. While $\eta_1 =1 $ is still optimal for the resonant mode, the optimal $\eta_2$ can vary significantly when $\omega_2$ is large with respect to the atomic transition. 
In Fig. \ref{threemode} we consider a 3-mode cavity (here we fix $\eta_3=1$ for the third mode). Again, we find that generally the optimal gauge is neither the dipole nor Coulomb gauge, and, most importantly, is mode-dependent.

\begin{figure*}[!ht]
\includegraphics[width=1\hsize]{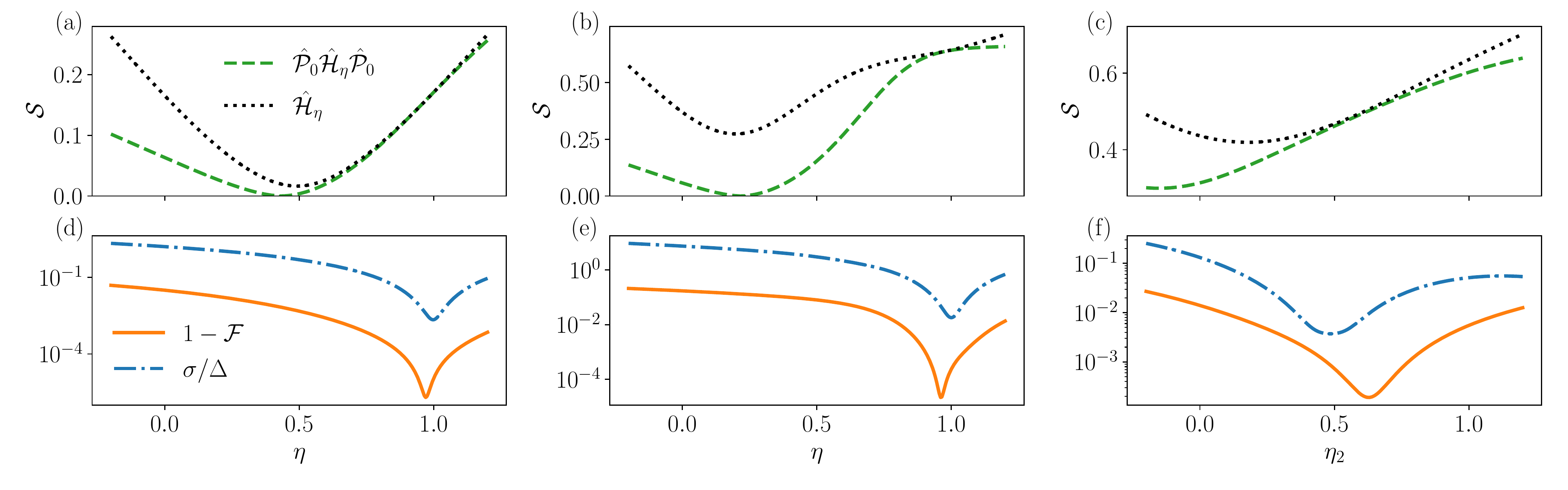}
\caption{The ground state entanglement entropy $\mathcal{S}$ of the full (black dotted) and of the truncated model (greed dashed) is plotted in the top panels (a), (b) and (c) as a function of the gauge parameter $\eta$. The spectral deviation $\sigma$ (dot-dashed) and the ground state fidelity $\mathcal{F}$ (solid) are plotted as a function of $\eta$ in the bottom panels (d), (e) and (f). Panels (a) and (d) are for a single-mode cavity with $\hbar \omega = \Delta $ and $g / \Delta=0.4$, while for panels (b) and (e) the coupling is $g / \Delta=1.2$. Panels (c) and (f) are for a two-mode cavity with $\hbar \omega_k = (1,20) \Delta $ and $g / \Delta=0.6$. The gauge of the first mode is fixed ($\eta_1 = 1$). 
\label{entanglement}
}
\end{figure*}

\subsection{Optimal gauge versus light-matter entanglement}
\label{entanglement_section}

The light-matter entanglement is not gauge invariant. This can be easily understood by noting that the transformation $ \hat{U}_{ \{  \eta \} } \neq \hat{U}_a  \otimes \hat{U}_p   $, i.e., it does not act separably on the atomic and photonic sectors. For the purpose of obtaining a reduced model, a fundamental question is whether the optimal gauge is somewhat related to the degree of entanglement. 
In this respect, it is certainly interesting to explore how the entanglement is modified when we project the full Hamiltonian to a truncated subspace. In the recent Ref. \cite{Ashida_2021}, it was discussed that low light-matter entanglement is desirable for an effective theory, since light-matter interaction can be handled more efficiently. To address this problem, here we investigate the gauge-dependent behavior of the ground state entanglement in truncated models, together with the spectral deviation $\sigma$ and ground-state fidelity ${\mathcal F}$ that we have already encountered in the previous sections.

The entanglement can be quantified by the entropy 
\begin{equation}
\mathcal{S}(\rho_{p})= - \text{Tr}[ \rho_{p} \log \rho_{p}]
\end{equation}
where $ \rho_{p} $ is the photonic reduced density matrix obtained by tracing out the atomic degrees of freedom. In the top panels of Fig. \ref{entanglement}, we report the entanglement entropy of the ground state for a single-mode (a-b) or two-mode (c) cavity.
In particular, we plot the exact results for the full Hamiltonian (black-dotted) and for the truncated model (green dashed) as a function of the gauge parameter $\eta$.
In the bottom panels (d-f), we display the corresponding values of the ground state infidelity $1 - {\mathcal F}$ (solid) and spectral deviation $\sigma$ (dot-dashed). Panels (a), (d) are for the case of a cavity mode resonant to the atomic transition frequency $\Delta$. In this situation, the dipole gauge is optimal (with respect to $\sigma$, and approximately for $1-\mathcal{F}$). However, the entanglement is minimal for a gauge parameter in-between the Coulomb and dipole gauges. 
Panels (b) and (e) are for the same configuration, but with a normalized coupling $g/\Delta$ three times larger than in panels (a) and (d).
In this case the difference between the entanglement of the full and the truncated models is increased, and the minimal entanglement is obtained for a different $\eta$.
Interestingly, for a single mode cavity, there is a value of $\eta$ where the ground state is not entangled in the truncated model. This point is identified with the vanishing of the non-rotating-wave terms so that the system is described \cite{stokes2019gauge} by the Jaynes-Cummings Hamiltonian. However, we note that for the full Hamiltonian, the entanglement entropy minimum is not zero. Moreover, the optimal gauge for the spectral deviation and ground state fidelity occurs for different values of $\eta$.

Panels (c) and (f) of Fig. \ref{entanglement} are for a two-mode cavity.  In this case, we fix $\eta_1=1$. The optimal $\eta_2$ for the spectral deviation and ground state fidelity is in-between the Coulomb and dipole gauges. Again, these optimal gauges do not correspond at all to a minimum of the ground state entanglement.
Finally, we note that the optimal gauge is not when the entanglement is high or low, but rather when the entropy \emph{difference} between the full and reduced models is minimal.

\section{Conclusions}
\label{conclusions}
In conclusion, we have shown that the optimal gauge for a truncated multi-mode cavity QED model is in general mode-dependent. Moreover, the optimal gauge can strongly depend on the observables of interest. In this work, we have focused on the spectral deviation (quantifying how the truncated model predicts the energy spectrum) and the ground state fidelity (quantifying how the truncated model can approximate the ground state wavefunction). While the gauge-dependence of a truncated cavity QED model is enhanced by strong light-matter interaction, we have shown that the degree of light-matter entanglement is not correlated to the optimal gauge. Indeed, the optimal gauge does not correspond at all to the minimum of light-matter entanglement. In our work, we have considered relatively simple cavity QED systems where it has been possible to compare the results of the full model to the truncated model, thus allowing us to rigorously determine the optimal gauges. An open  problem that represents a fascinating perspective for the future is the search for some criteria that allows to systematically determine the optimal gauges for arbitrary models. This is an interesting issue that is certainly crucial to tackle more complex cavity QED systems with a larger number of atomic degrees of freedom and cavity modes.  

\acknowledgements
{We thank N. Mehta for helpful discussions. We acknowledge support from the Israeli Council for Higher Education - VATAT, from FET FLAGSHIP Project PhoQuS (grant agreement ID no.820392) and from the French agency ANR through the project NOMOS (ANR-18-CE24-0026), TRIANGLE (ANR-20-CE47-0011) and CaVdW (ANR-21-CE30-0056-01).}

\bibliography{gaugerefs}

\end{document}